\documentclass[lettersize,journal]{IEEEtran}

\usepackage[switch,columnwise]{lineno}
\usepackage{lipsum}
\usepackage{threeparttable}
\usepackage[dvipsnames,svgnames,x11names,table]{xcolor}
\usepackage{siunitx}
\usepackage{amsmath}
\usepackage{amssymb}
\usepackage{physics}
\usepackage{xfrac}
\usepackage{nicefrac}
\usepackage{bm}
\usepackage{mathtools}
\usepackage{multirow}
\usepackage[utf8]{inputenc}
\usepackage{booktabs}
\usepackage{arydshln}
\usepackage{array}
\usepackage{chemformula}
\usepackage{amsmath,amsfonts}
\usepackage{algorithmic}
\usepackage{algorithm}
\usepackage{array}
\usepackage[caption=false,font=normalsize,labelfont=sf,textfont=sf]{subfig}
\usepackage{textcomp}
\usepackage{stfloats}
\usepackage{url}
\usepackage{verbatim}
\usepackage{graphicx}
\usepackage{cite}
\usepackage{microtype}
\usepackage[]{hyperref}
\hypersetup{
	pdfauthor={Thomas Daunizeau},
	pdfcreator={Thomas Daunizeau},
	pdfproducer={Thomas Daunizeau},
	colorlinks,
	citecolor=black,
	filecolor=black,
	linkcolor=black,
	urlcolor=DarkBlue
}

\NewDocumentCommand{\evalat}{sO{\big}mm}{\IfBooleanTF{#1}{\mleft. #3 \mright|_{#4}}{#3#2|_{#4}}}
\newcommand{\pdm}[6]{\frac{\partial^#1{#2}}{\partial{#4}^#3 \partial{#6}^#5}}

\begin{document}

% -----
% TITLE
% -----
\title{Optimized Sandwich and Topological Structures\\for Enhanced Haptic Transparency}

% -----------------------
% AUTHORS AND AFFILIATION
% -----------------------
\author{Thomas Daunizeau$^{1}$, Sinan Haliyo$^{1}$, and Vincent Hayward$^{1}$
\thanks{$^{1}$ Thomas Daunizeau, Sinan Haliyo, and Vincent Hayward (deceased) are with Sorbonne Universit\'{e}, Institut des Syst\`{e}mes Intelligents et de Robotique, ISIR, 75005 Paris, France. \\ {\tt\small thomas.daunizeau@sorbonne-universite.fr}}
}

% ---------------
% HEADER & FOOTER
% ---------------
\markboth{IEEE TRANSACTIONS ON HAPTICS,~Vol.~XX, No.~X, MONTH~YEAR}
{Shell \MakeLowercase{\textit{et al.}}: A Sample Article Using IEEEtran.cls for IEEE Journals}
%\IEEEpubid{XXXX--XXXX/XX\$XX.XX~\copyright~YEAR IEEE}

\maketitle

% ------------------------
% ABSTRACT (180/200 WORDS)
% ------------------------
\begin{abstract}

Humans rely on multimodal perception to form representations of the world. This implies that environmental stimuli must remain consistent and predictable throughout their journey to our sensory organs. When it comes to vision, electromagnetic waves are minimally affected when passing through air or glass treated for chromatic aberrations. Similar conclusions can be drawn for hearing and acoustic waves. However, tools that propagate elastic waves to our cutaneous afferents tend to color tactual perception due to parasitic mechanical attributes such as resonances and inertia. These issues are often overlooked, despite their critical importance for haptic devices that aim to faithfully render or record tactile interactions. Here, we investigate how to optimize this mechanical transmission with sandwich structures made from rigid, lightweight carbon fiber sheets arranged around a 3D-printed lattice core. Through a comprehensive parametric evaluation, we demonstrate how this design paradigm provides superior haptic transparency, regardless of the lattice types. Drawing an analogy with topology optimization, our solution approaches a foreseeable technological limit. It offers a practical way to create high-fidelity haptic interfaces, opening new avenues for research on tool-mediated interactions.

\end{abstract}

% --------
% KEYWORDS
% --------

% ------------
% INTRODUCTION
% ------------
\section{Introduction}

\IEEEPARstart{T}{he} use of tools extends somatosensory processing beyond the physical self~\cite{Loomis-92, MillerEtAl-19}. For example, visually impaired persons can safely navigate with a white cane, while dentists are able to detect cavities with a sickle probe. Even mundane objects, such as a pen, can reveal the subtle grain of a piece of paper. Tools benefit from being made out of materials orders of magnitude stiffer than skin. Wood for instance, has a Young's modulus $E\approx \SI{10}{\giga\pascal}$ along grain opposed to, say $E\approx \SI{10}{\kilo\pascal}$ for the skin~\cite{vanKuilenburgEtAl-13}. In turn, its rigid interfacial asperities magnifies frictional noise induced by Coulombic interactions, even more so if the tool is sharp. Short transients generated upon impacts and interlocking friction create vivid vibrotactile sensations, which humans are exquisitely apt to perceive up to about $\SI{1}{\kilo\hertz}$~\cite{JohanssonFlanagan-09}. More generally, tool-mediated interactions convey a wide range of tactile cues that help discriminate softness~\cite{LaMotte-00}, roughness~\cite{KlatzkyLederman-99}, vertical distance~\cite{ChanTurvey-91}, and position~\cite{SaigEtAl-12}.

Recent findings indicate how humans can localize impacts on a handheld wooden rod, as accurately as if it was their own arm~\cite{MillerEtAl-18}. In substance, the somatosensory cortex efficiently maps a stereotypical modal response to the impact location. However, this ability vanishes for overly soft rods. This underscores perceptual priors on structural dynamics and the critical role of material properties on the way humans wield tools.

\begin{figure}[!t]
\centering
\includegraphics[width=88mm]{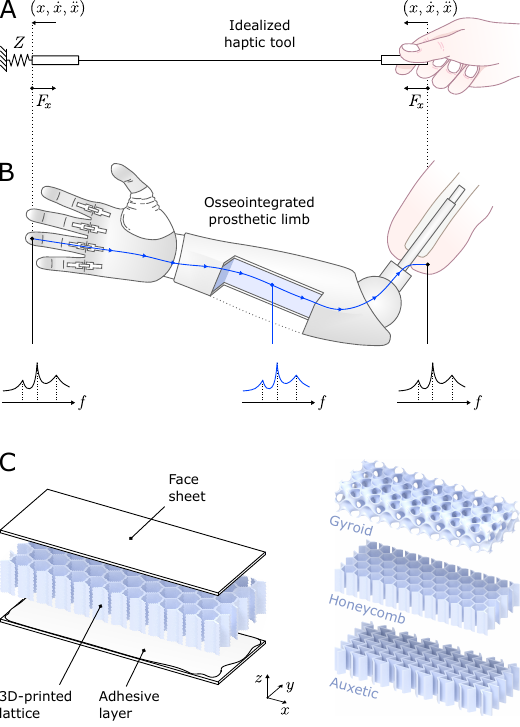}
\caption{(A) An ideal haptic tool as a massless and infinitely stiff rod, inspired by~\cite{HaywardMaclean-07}. It would faithfully render the impedance of the environment, $Z$, either real or virtual. (B) An osseointegrated prosthesis as an example of a real-world haptic tool. Drawing inspired by~\cite{FarinaEtAl-21}. As revealed by the cut-out, it could be made out of an optimal haptic medium that propagates artifact-free stimuli, from a distal contact up to the residual limb. (C) Exploded view of an optimal haptic medium, introduced here as a sandwich structure. Its 3D-printed core can be one of the following lattice stuctures: gyroid, honeycomb, or auxetic.}
\label{fig:design_paradigm}
\end{figure}

An ideal haptic tool has no mass and an infinite stiffness~\cite{HaywardMaclean-07}. If equipped with such a tool, one would feel the environment unaltered, as if the tool did not exist. Contact forces at both its extremities would be of equal magnitude and opposite sign at any time (see Fig.~\ref{fig:design_paradigm}.A). Such ability to render undistorted stimuli is commonly referred to as haptic transparency~\cite{LawrenceChapel-94}. While not all tools can be reduced to a massless rigid stick, it is sufficient to meet the limits of human performance. Accordingly, haptic tools should have a stiffness of at least \SI[inter-unit-product = \!\cdot\!]{40}{\newton\per\milli\meter}\!, a mass below \SI{5}{\gram}, and a mechanical bandwidth of at least \SI{1}{\kilo\hertz}~\cite{TanEtAl-94, ColgateBrown-94, MilletEtAl-09}. In practice, structures achieving the target bandwidth with minimal mass are also sufficiently stiff due to fundamental principles of continuum mechanics and material properties. Yet, these target figures are often overlooked, despite their role in both kinesthetic and vibrotactile feedback.

This is the case for prosthetic limbs, most of which violate the rules of transparency due to poor weight distribution, significant inertia, and suboptimal rigidity. Prostheses in clinical use lack a sense of touch, setting up a major challenge. A newly discovered feedback modality, called ``osseoperception'' offers promising results. Unlike most prostheses, which are mounted on sockets, osseointegrated artificial limbs are anchored directly to the bone, as shown in Fig.~\ref{fig:design_paradigm}.B. The improved mechanical coupling creates a strong vibratory pathway. This can trigger cortical activity~\cite{Habre-HallageEtAl-12} and yield marked improvements on the detection threshold of vibrotactile stimuli around \SI{250}{\hertz}~\cite{HaggstromEtAl-13}. Bypassing the prosthesis and stimulating directly the abutment screw revealed that elastic waves up to \SI{6}{\kilo\hertz} not only radiated to nearby somatosensory neurons, but could also reach the organ of Corti, thus involving both touch and hearing~\cite{ClementeEtAl-17}. In \mbox{practise}, the bandwidth of osseoperception appears to be capped by the prosthetic structure itself, which, if carefully designed, could become a perfectly viable haptic tool.

% ------
% FIGURE
% ------

Handheld tools and prosthetic limbs are designed to be used in-situ. Alternatively, there is an ever-growing need to extend our haptic reach to remote and virtual environments. Over the past decades, this has catalyzed the creation of a broad ecosystem of force feedback devices~\cite{SeifiEtAl-19}. Their frequency response is yet rarely discussed, let alone measured, despite its fundamental importance. For example, rendering short transients is essential to the perception of stiff virtual walls~\cite{KuchenbeckerEtAl-06}. Some commercial devices have a bandwidth as low as \SI{20}{\hertz}~\cite{HaywardEtAl-98}, a far cry from established requirements. Parallel kinematics is a common solution to increase stiffness and reduce moving mass by grounding the actuators. In this way, significantly higher bandwidths have been reported, e.g. \SI{175}{\hertz}~\cite{LeeEtAl-00}. Refined devices with limited workspace can even reach \SI{300}{\hertz}, after which sharp metallic resonances occur~\cite{CampionEtAl-05}.

In his article on geometrical axioms~\cite{Helmholtz-76}, Helmholtz emphasized that ``all our geometrical measurements depend on our instruments being really, as we consider them, invariable in form''. This principle applies equally to tactile sensing, which faces the same issues undermining haptic rendering. Bio-tribometers are designed to record artifact-free frictional forces. Instead, most suffer from normal modes between \SI{300}{\hertz} and \SI{600}{\hertz}~\cite{BochereauEtAl-17,GrigoriiEtAl-17,BernardEtAl-19}, although reconstruction filters can, in some cases, further extend the usable bandwidth up to \SI{1.3}{\kilo\hertz}~\cite{BernardEtAl-19}.
 
Haptic engineers face the arduous task of making transparent devices whose structures are currently limited by outdated design and manufacturing approaches. Advances in materials science have already paved the way to novel means of actuation and sensing for haptics~\cite{BiswasVisell-19}. Along similar lines, this paper explores how emerging material technologies could be leveraged in a novel design paradigm to assist haptic engineers in creating highly transparent tools. Please note that a broad definition of tools has been adopted here, including passive probes, active force feedback devices, and measuring instruments.

% ---------------
% DESIGN PARADIGM
% ---------------
\section{Design Paradigm}

Bending is the bottleneck of nearly all haptic tools. In fact, continuum mechanics predicts the first normal mode of elongated structures to be flexural. This longstanding issue was addressed in 1820 by Duleau who first theorized a lightweight frame with faces spaced apart from the neutral axis~\cite{Duleau-20}. It achieved an unprecedented bending stiffness-to-weight ratio. This would later be known as a ``sandwich material''. Although widespread in the aerospace industry, it has not permeated yet the field of haptics. The sandwich architecture in Fig.~\ref{fig:design_paradigm}.C is the backbone of the novel design paradigm introduced here.

\subsection{3D-printed Cellular Cores}

The core plays arguably a critical role in sandwich panels, that is to prevent relative motion between face sheets without adding unnecessary mass. Hence why it must be of low-density and have both high shear and compression moduli. As such, cores are typically made from end-grain balsa, cellular foams, or aramid honeycombs. Progress in additive manufacturing has enabled thermoplastic cellular cores of shapes previously beyond reach~\cite{PengEtAl-21}. Unlike conventional cores with limited design options, these 3D-printed lattices can be arbitrarily tailored to a desired goal, e.g. maximizing haptic transparency. Numerous quasi-static and failure analyses have been reported on lattices~\cite{LiWang-17,YazdaniSarvestaniEtAl-18,PengEtAl-21}. However, experimental evidence of their dynamic responses is still lacking. Due to their complex geometries and ill-defined material properties, the design of 3D-printed cores remains much of an empirical process. The following three lattices (see Fig.~\ref{fig:design_paradigm}.C), selected on the basis of educated guesses, were to be investigated experimentally.\\

\begin{itemize}

\item[$\bullet$] The first lattice is a minimal surface, that is a surface which minimizes its area while spanning a given boundary. In 1768, Lagrange laid the theoretical foundations of minimal surfaces which were first described empirically by Plateau in 1873~\cite{Plateau-73}. By dipping wire frames in soap solution, he showed that soap films, stretched by surfactants, formed surfaces of minimal area. It was much later, in 1970, that Schoen described mathematically the triply periodic gyroid structure~\cite{Schoen-70}, shown in Fig.~\ref{fig:design_paradigm}.C. This minimal surface lattice benefits from quasi-isotropic mechanical properties at a minimum weight. It is approximated by,
\begin{equation}\label{eq:gyroid}
\cos(x)\sin(y)+\cos(y)\sin(z)+\cos(z)\sin(x)=0
\, .
\end{equation}
\item[$\bullet$] The second lattice is the widely used honeycomb structure, inspired by beehives. It benefits from the highest shear stiffness-to-weight ratio of all available cores. This is because a regular hexagonal grid is the geometry of lowest perimeter to tile the plan, as demonstrated by Hales who solved the honeycomb conjecture~\cite{Hales-01}.
\\
\item[$\bullet$] The third lattice is an auxetic structure, i.e. with a negative Poisson's ratio~\cite{Lakes-87}. This unusual property leads to unique acoustic absorption capabilities~\cite{HowellEtAl-94}, which makes it a good candidate to dampen sharp structural resonances. The geometry shown in Fig.~\ref{fig:design_paradigm}.C gains its auxetic nature from re-entrant honeycomb ribs tilted at an angle $\theta$. It yields a Poisson's ratio, $\nu_x$, defined by equation~\eqref{eq:auxetic} along the $x$-axis~\cite{WanEtAl-04}. The reciprocal in-plane ratio, $\nu_y$, must satisfy $\nu_x \nu_y=1$. An angle $\theta=\SI{50}{\degree}$\!, restricted by manufacturing constraints, gave $\nu_x \approx -0.39$ and $\nu_y \approx -2.56$.
\begin{equation}\label{eq:auxetic}
\nu_x=\frac{\cos^2(\theta)-\cos(\theta)}{\sin^2(\theta)}
\, .
\end{equation}

\end{itemize}

% ------
% FIGURE
% ------
\begin{figure}[!b]
\centering
\includegraphics[width=88mm]{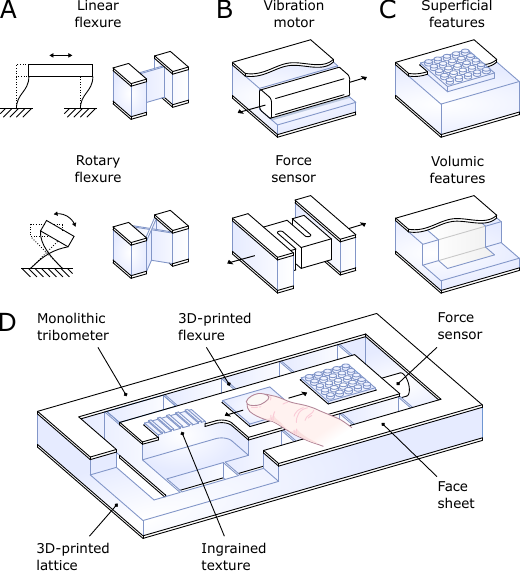}
\caption{(A) Example of a functionalized core with integrated flexures in either translation or rotation. (B) Example of seamless mounting points for embedded actuators and sensors. (C) Example of superficial or volumic tactile features embedded in the core. (D) Example of a combination of several functional blocks to create a monolithic, highly transparent tribometer that includes a flexure-guided linear stage, a force sensor, and customizable ingrained textures.}
\label{fig:functional_blocks}
\end{figure}

\subsection{Functional Blocks}

An added benefit of 3D-printed cores is the possibility to embed functional blocks within a single monolithic part. For example, flexures can be added to set in motion different sections of the lattice, either in rotation or translation as illustrated in Fig.~\ref{fig:functional_blocks}.A. Despite being manufactured in layers, this type of frictionless, backlash-free linkages can still maintain great positioning accuracy with a drift reported below \SI{20}{\micro\meter} per week~\cite{SharkeyEtAl-16}. The lattice could also be altered to provide seamless mounting points for tactile actuators and sensors, as shown in Fig.~\ref{fig:functional_blocks}.B. Areas of the 3D-printed structure could also sprout through the face sheets and provide tailored tactile cues, as shown in Fig.~\ref{fig:functional_blocks}.C. Textures could be ingrained~\cite{OuEtAl-16,IonEtAl-18} and the lattice could even be altered within the volume to modulate the perception of softness~\cite{MiyoshiEtAl-21}. Taken altogether, these functional blocks can be arranged, for instance, to make a high-fidelity tribometer, as depicted in Fig.~\ref{fig:functional_blocks}.D. The enticing applications of functional blocks are beyond the scope of this study and were not further investigated.

% ------------------------------
% EFFECTIVE SANDWICH BEAM MODELS
% ------------------------------
\section{Effective Sandwich Beam Models}

The proposed sandwich architecture must be optimized to meet the desired transparency requirements. Its bandwidth is set by the first resonance mode. The following modal analysis was conducted to identify relevant design factors. For the sake of generalizability, it was limited to rectangular beams with dimensions representative of common haptic tools (see~\ref{sec:dimensional_optimization}).

% ------
% FIGURE
% ------
\begin{figure}[!h]
\centering
\includegraphics[width=88mm]{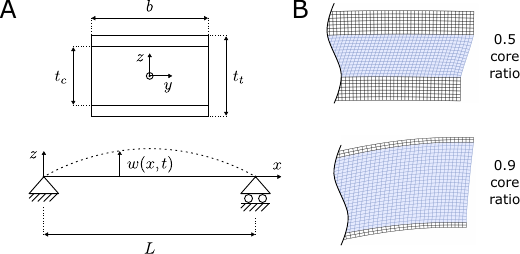}
\caption{Effective sandwich beam models. (A) Schematic of the analytical model with simply supported extremities. The dashed line shows the transverse displacement of the neutral axis. (B) Numerical FEA model, represented with close-up views of the extremity of a free-free beam of thickness \SI{12.6}{\milli\meter} with 0.5 and 0.9 core ratios. Displacements are magnified for visualization.}
\label{fig:models}
\end{figure}

\subsection{Sandwich Beam Theory}

The sandwich structure was modeled as a thick beam of length $L$, width $b$, and total thickness $t_t$, as shown in Fig.~\ref{fig:models}.A. It consisted of a symmetrical lay-up of two face sheets of same thickness, $t_s$, arranged around a core of thickness $t_c$. The proportion, $r$, of core material was defined as $t_c/t_t$. The layers were assumed to be perfectly bonded together, ensuring continuity of motion at their interfaces.

The lattices were designed with sufficient cells to yield near homogeneous properties at a macroscopic level. Based on an effective medium approximation (EMA), their intricate inner structure could therefore be simplified to that of a single block endowed with effective properties. The principal load the core had to resist was shear along the $x$-axis to avoid faces sliding past each other. Accordingly, it was to follow a linear elastic behavior characterized by a Young's modulus $E_c$, a shear modulus $G_c$, and a Poisson's ratio $\nu_c$. Lattices presented in Fig.~\ref{fig:design_paradigm}.C are inherently orthotropic, which invalidates the usual relation $G_c=E_c/2(1+\nu_c)$. Nonetheless, linearity still prevails under the small strain assumption, hence $G_c=\varsigma \, E_c$, with $\varsigma$ a function of the lattice geometry. This is true for honeycomb lattices, among others~\cite{GibsonEtAl-82}. Both face sheets were assumed to be homogeneous, quasi-isotropic, and linear elastic, with a Young's modulus, $E_s$, defined along the $x$-axis, a Poisson's ratio $\nu_s$, and a shear modulus, $G_s$, defined as $G_s=E_s/2(1+\nu_s)$. The beam was assumed to undergo bending along the $y$-axis. Its deformation was described by the Timoshenko-Ehrenfest theory, which accounted for both rotary inertia and shear deformation. To fit within this framework, the beam was given effective properties indicated by an asterisk. The effective mass per unit length, $\rho^*$, was defined as,
\begin{equation}
\label{eq:equivalent_density}
\rho^* = 2 \rho_s b\,t_s + \rho_c b\,t_c
\, ,
\end{equation}
which could be factorized into, 
\begin{equation}\label{eq:equivalent_density_factorized}
\rho^* = \rho_s b\,t_t\, [1+r(\beta-1)]
\, ,
\end{equation}
with the dimensionless ratio, $\beta\!=\!\rho_c/\rho_s$, of the core and face sheets densities, $\rho_c$ and $\!\rho_s$, respectively. The flexural rigidity of the laminate around the $y$-axis, $D^*$\!, is the sum of contributions from the face sheets, $D_s$, the core, $D_c$, and an additional term, $D_0$, that captures the so-called sandwich effect~\cite{Duleau-20}, such as,
\begin{equation}\label{eq:equivalent_rigidity}
D^* = 2D_s + D_0 + D_c
\, ,
\end{equation}
\begin{equation}\label{eq:equivalent_rigidity_expanded}
D^* = \frac{E_s b\,{t_s}^3}{6} + \frac{E_s b\,t_s (t_c+t_s)^2}{2} + \frac{E_c\, b\,{t_c}^3}{12}
\, ,
\end{equation}
which could be factorized into, 
\begin{equation}\label{eq:equivalent_rigidity_factorized}
D^* = E_s b\,{t_t}^{\!3} \bigg[\frac{(1-r)^3}{48} + \frac{(1-r)(1+r)^2}{16} + \frac{\alpha r^3}{12} \bigg] 
\, ,
\end{equation}
with the dimensionless ratio $\alpha=E_c/E_s$. Similarly, the rotary inertia, $R^*$\!, was defined as,
\begin{equation}\label{eq:equivalent_inertia_factorized}
R^* = \rho_s b\,{t_t}^{\!3} \bigg[\frac{(1-r)^3}{48} + \frac{(1-r)(1+r)^2}{16} + \frac{\beta r^3}{12} \bigg] 
\, .
\end{equation}
The effective shear stiffness, $S^*$\!, was given by,
\begin{equation}\label{eq:equivalent_shear_stiffness}
S^* = \frac{G_c\,\kappa\,b\,(t_c+t_s)^2}{t_c}
\, ,
\end{equation}
with $\kappa \approx 5/6$, the shear coefficient of a rectangular cross section. Equation~\eqref{eq:equivalent_shear_stiffness} could be factorized into,
\begin{equation}\label{eq:equivalent_shear_stiffness_factorized}
S^* = \frac{E_s\kappa\,\varsigma\, b\,t_t\, (1+r)^2}{4r}
\, .
\end{equation}
The natural response of the sandwich beam was expressed by the following wave equation, describing the transverse displacement $w$ along the $z$-axis such as,
\begin{equation}\label{eq:effective_wave_timoshenko}
D^*\!\pdv[4]{w}{x} \;\!\! + \;\!\! \rho^* \! \pdv[2]{w}{t} \;\!\! - \;\!\! \frac{R^* \! S^* \!\! + \! D^* \! \! \rho^*}{S^*} \pdm 4{w} 2x 2t \;\!\! + \;\!\! \frac{R^* \!\! \rho^*}{S^*}\pdv[4]{w}{t} \! = \! 0
\, .
\end{equation}
In sandwich materials, the contribution of rotary inertia is much smaller than that of bending stiffness, hence,
\begin{equation}\label{eq:inegality_rotary_bending}
\frac{R^*}{D^*} \ll \frac{\rho^*}{S^*} 
\, .
\end{equation}
The wave equation~\eqref{eq:effective_wave_timoshenko} could thus be reduced to,
\begin{equation}\label{eq:effective_wave_timoshenko_reduced}
\pdv[4]{w}{x} + \frac{\rho^*}{D^*} \pdv[2]{w}{t} - \frac{\rho^*}{S^*} \pdm 4{w} 2x 2t = 0
\, .
\end{equation}
This eigenvalue problem was to be solved under the assumption of simply supported extremities,
\begin{equation}\label{eq:boundary_simply_supported}
\evalat[]{w}{x=0} = \evalat[]{w}{x=L} = 0
\, .
\end{equation}
Other types of boundaries lack closed-form expressions. They require numerical solutions instead, which prevent analytical optimization and hinder physical insights. Nonetheless, a sense of generality was maintained as the governing principles found in the following analysis still hold for alternative boundary conditions, e.g. free-free.
A variational method was used to find the solution of the wave equation~\eqref{eq:effective_wave_timoshenko_reduced}. The action functional, $F$, was defined as,
\begin{equation}\label{eq:functional_F}
F = \int_{0}^{\,2\pi/\omega} \!\! \mathcal{L}\,\, dt
\, ,
\end{equation}
where $\omega$ is the harmonic frequency and $\mathcal{L}$ is the Lagrangian, defined as, $\mathcal{L}=T\!-\!U$, with $T$ and $U$ the kinetic and potential energies, respectively. Hamilton's principle states that the solution of equation~\eqref{eq:effective_wave_timoshenko_reduced} is a stationary point of $F$. This was solved using Rayleigh-Ritz's method. The resonance frequency of the $n$\textsuperscript{th} flexural mode is $\omega_n$, defined as,
\begin{equation}\label{eq:resonance_sandwich}
\omega_n = \sqrt{\frac{(n\pi/L)^4 D^*\!/\rho^*}{1+(n\pi/L)^2 D^*\!/S^*}}
\, ,
\end{equation}
which can be rewritten as,
\begin{equation}\label{eq:resonance_sandwich_factorized}
\omega_n = \omega_{s}\eta
\, ,
\end{equation}
where $\omega_{s}$ represents the limit case of an homogeneous beam that would be made entirely of face sheet material and not subjected to shear, such as,
\begin{equation}\label{eq:resonance_frequency_face_sheet}
\omega_{s} = \Big(\frac{n\pi}{L}\Big)^{\!2} \sqrt{\frac{{t_t}^{\!2} E_s}{12\,\rho_s}} 
\, ,
\end{equation}
and where $\eta$ is a frequency gain capturing the effects of a sandwich layout on the natural response, defined as,
\begin{equation}\label{eq:factor_total}
\eta = \sqrt{\frac{\chi}{1+\phi\,(n\pi t_t /L)^2}}
\, .
\end{equation}
The frequency gain, $\eta$, is a function of $\chi$, primarily due to bending, such as,
\begin{equation}\label{eq:factor_bending}
\chi = \frac{1}{1+r(\beta-1)}\bigg[\frac{(1-r)^3}{4} \! + \! \frac{3(1-r)(1+r)^2}{4} \! + \alpha r^3 \bigg] 
\, ,
\end{equation}
and a function of $\phi$, primarily due to shear, such as,
\begin{equation}\label{eq:factor_shear}
\phi = \frac{r}{\alpha\,\kappa\,\varsigma\, (1+r)^2}\bigg[\frac{(1-r)^3}{12} \! + \! \frac{(1-r)(1+r)^2}{4} \! + \! \frac{\alpha r^3}{3} \bigg]
\, .
\end{equation}
The first resonance frequency, $\omega_1$, given by equation~\eqref{eq:resonance_sandwich_factorized} for $n=1$, will limit the bandwidth. This assumes bending along the $y$-axis. If one was to consider bending along the $z$-axis instead, it would result in a higher resonance frequency. We made similar observations for torsional modes. In other words, this study judiciously addresses the worst case scenario.

\subsection{Finite Element Analysis}

A numerical model was developed to complement analytical investigations. Its purpose was twofold. First, it provided a verification that took into account nonlinear geometric effects not captured by beam theory. Second, it offered a solution for free-free boundary conditions, which are otherwise unattainable analytically, yet of great practical importance. Unsupported boundaries are indeed the simplest to replicate accurately, given that any rigid bond or clamp, even designed with utmost care, would introduce discrepancies. Accordingly, a finite element analysis (FEA) was devised on COMSOL Multiphysics 5.5. The beam was approximated by a 2D geometry under the plane stress assumption ($\sigma_{yy}\!=0$). This simplified representation enabled an extremely fine meshing of the thin face sheets without prohibitive computational cost. Accordingly, the adaptive mesh included elements with a characteristic length down to $\SI{0.1}{\milli\meter}$ (see Fig.~\ref{fig:models}.B). Both analytical and numerical models laid the groundwork for the subsequent optimization.

% ------------
% OPTIMIZATION
% ------------
\section{Optimization}

% ------
% FIGURE
% ------
\begin{figure}[!b]
\centering
\includegraphics[width=88mm]{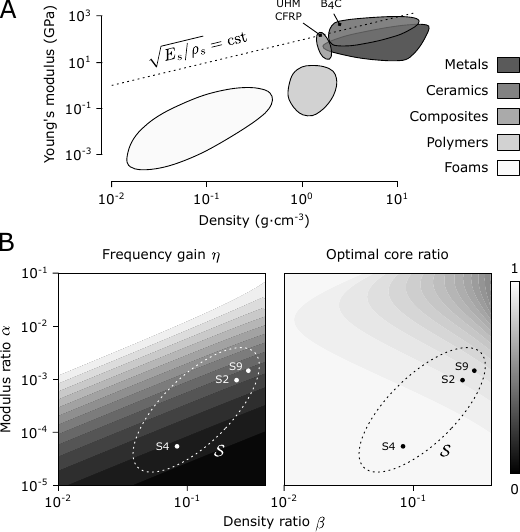}
\caption{(A) Material property chart known as a Ashby plot. It shows the relationship between Young's modulus and density for common families of materials. The dashed line indicates a constant speed of sound of $10^4$\,\SI[inter-unit-product = \!\cdot\!]{}{\meter\per\second}\!. (B) Theoretical frequency gain caused by the sandwich effect, given as a function of the core material properties and computed for an optimal core ratio. Points S2, S4, and S9 represent the honeycomb cores used in the experiment.}
\label{fig:material_optim}
\end{figure}

\subsection{Face Sheet Material Optimization}

The target optimal haptic medium calls for a bandwidth of at least $\SI{1}{\kilo\hertz}$ and a minimum mass. The goal to prioritize depends on the end-use application and is thus left to the reader's judgment. To enhance the bandwidth, the first normal mode $\omega_1$ should be as high as possible. An established result of beam theory is that natural frequencies are proportional to the material index $\sqrt{E/\rho}$. This is why steel rings at a higher pitch than softwoods upon hammering. Interestingly, similar findings hold for a sandwich beam as its normal modes are also function of the same index, albeit applied to the face sheets, $\sqrt{E_s/\rho_s}$, as shown in equation~\eqref{eq:resonance_frequency_face_sheet}. Ashby devised a well-known graphical method to select materials that maximize this performance index~\cite{Ashby-99}. Such material selection chart, also known as an Ashby plot, is presented in Fig.~\ref{fig:material_optim}.A. Materials lying on the same dashed line share the same index. To establish a shortlist of suitable materials, this line was translated across the chart until only a small subset was left above it. Both families of carbon fiber reinforced polymers (CFRP) and technical ceramics were isolated accordingly. Among composites, ultra-high modulus (UHM) panels can reach extreme index values of $\SI[inter-unit-product = \cdot]{13764}{\meter\per\second}$\!, more than twice that of aluminium or steel, $\SI[inter-unit-product = \cdot]{5327}{\meter\per\second}$\!. These data were obtained experimentally from a three-point bending test. These are, in fact, the speeds of sound in solids. Some ceramics can keep up to similar index levels, i.e. boron carbide (\ch{B4C}) at $\SI[inter-unit-product = \!\cdot\!]{13511}{\meter\per\second}$\!, but cannot be manufactured in thin sheets. Additionally, CFRP have a competitive low density of only $\SI[inter-unit-product = \!\cdot\!]{1.55}{\gram\per\centi\meter\cubed}$. An added benefit of fibrous composites is their intrinsic structural damping~\cite{AdamsEtAl-69}. This is the result of energy dissipation from a myriad of microscopic frictional contacts at the fibers periphery, coupled with the viscoelasticity of the polymer matrix. Sufficient damping could tame otherwise sharp resonances and increase bandwidth accordingly.

Therefore, fabricating face sheets from UHM-CFRP would unambiguously outperform any other solution aimed at minimizing mass and increasing bandwidth. 

\subsection{Core Material Optimization}

Finding an optimal core material, if any, is not as straightforward as it is for the face sheets. The latter thus served as a reference from which core properties were defined with the aforementioned ratios $\alpha$ and $\beta$. To enhance the bandwidth, the frequency gain, $\eta \,(\alpha,\beta)$, needs to be maximized. Therefore, it is necessary, but not sufficient, to have critical points such as,\begin{equation}\label{eq:maximum_eta}
\bigg[\pdv[]{\eta}{\alpha} \,, \pdv[]{\eta}{\beta}\bigg] = (0,0)
\, .
\end{equation}
This can equally, and more easily, be applied to $\eta^2$. However, $\partial \eta^2 / \partial \alpha > 0$ and $\partial \eta^2 / \partial \beta < 0 \, \forall \, (\alpha,\beta) \in \mathbb{R}^2_{>0}$. Furthermore, since $\eta$ is a strictly monotonic function of both $\alpha$ and $\beta$, there is no theoretically optimal core material. Instead, the frequency gain, $\eta$, was evaluated for a broad range of possible values of $\alpha$ and $\beta$, as shown in Fig.~\ref{fig:material_optim}.B. The 3D-printer available (S3, Ultimaker) limited the space of possible material properties to a subspace, denoted by $\mathcal{S}$. Its boundaries, delineated in Fig.~\ref{fig:material_optim}.B, encompass the 3D-printed honeycomb lattices introduced in this paradigm, labeled S2, S4, and S9. The remaining lattices were not included owing to a lack of effective core properties in the literature. The frequency gain, $\eta$, is below unity inside the subset $\mathcal{S}$. This indicates that a sandwich beam would have a lower bandwidth compared to a beam made entirely of face material. Although it may seem counterintuitive, the benefit of the sandwich layout actually lies in its reduced inertial mass rather than an increase in bandwidth. As previously stated, balancing both factors hinges on the end-use application. This can be achieved by selecting a core closer to either S2 and S9 or to S4, for example.

Theoretical results in the left panel of Fig.~\ref{fig:material_optim}.B were obtained independently of the core ratio, $r$, to account only for material contribution. This was achieved by finding an optimal core ratio, $r$, for each pair $(\alpha,\beta)$, as shown on the right panel. This one-dimensional optimization problem was solved numerically using golden section search. For low-density lattices (e.g. S4), the core ratio is close to unity, which could lead to excessively thin face sheets that are challenging to fabricate.

% ------
% FIGURE
% ------
\begin{figure*}[!t]
\centering
\includegraphics[width=181mm]{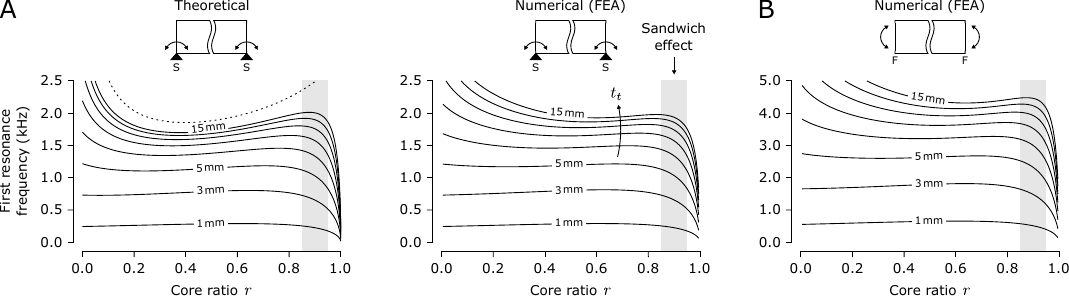}
\caption{Estimates of the first resonance frequency of a sandwich beam as a function of the core ratio, $r$, and total thickness, $t_t$, from 1 to \SI{15}{\milli\meter}. Results were computed with material properties from sample S2. (A) Theoretical and numerical results for simply supported boundary conditions. The dashed line depicts $f_t=\omega_{t}/2\pi$. (B) FEA evaluation for free-free boundary conditions.}
\label{fig:dimensional_optim}
\end{figure*}

\subsection{Dimensional Optimization}\label{sec:dimensional_optimization}

Guidelines for optimal material selection were derived. To find optimal dimensions, the problem was simplified by fixing arbitrarily the length, $L=\SI{160}{\milli\meter}$, and the width, $b=\SI{30}{\milli\meter}$. This strikes a good balance between ease of prototyping and a form factor suited to common haptic applications. It should be noted that the width, $b$, has no effect other than modifying the inertial mass in direct proportion. The first resonance frequency, $\omega_1$, is presented in Fig.~\ref{fig:dimensional_optim}.A and Fig.~\ref{fig:dimensional_optim}.B as a function of the remaining dimensions, $r$ and $t_t$, for both simply supported and free-free boundaries. The theoretical model is in good agreement with the finite element analysis, thereby further confirming previous findings. While these results were computed for sample S2, similar conclusions could be drawn for other samples. The resonance frequency of the simply supported beam is lower, around half as low as that of the free-free beam, as a result of forced nodal spacing. When the form factor $t_t/L$ is such that $\phi\,(n\pi t_t /L)^2\gg 1$, the resonance frequency tends towards a threshold, $\omega_t$, beyond which any increase in thickness, $t_t$, has a negligible effect. This threshold is given by,
\begin{equation}\label{eq:resonance_threshold}
\omega_{t} = \frac{n\pi}{L} \sqrt{\frac{E_s\chi}{12\,\rho_s\,\phi}} 
\, ,
\end{equation}
and illustrated by the dashed line in Fig.~\ref{fig:dimensional_optim}.A. Based on this observation and taking into account manufacturing constraints, the total thickness, $t_t$, was set to \SI{12.6}{\milli\meter}. It yields normalized cuboid samples ($160\! \times \!30\! \times \!\SI{12.6}{\milli\meter\cubed}$), ideal for benchmarking haptic transparency.

The fraction of core material that maximizes transparency remains to be found. Both boundary conditions indicate that the so-called sandwich effect is more pronounced for laminates with thin outer faces. This region, shaded in Fig.~\ref{fig:dimensional_optim}.A and Fig.~\ref{fig:dimensional_optim}.B, is characterized by a resonance frequency with a local maximum and a corresponding reduction in mass. The optimal core ratio, when $\partial \eta / \partial r = 0$, is nearly independent of the beam thickness. Based on these findings, the core ratio, $r$, was set to $0.90$, as illustrated in Fig.~\ref{fig:models}.B. Accurate thickness control during prototyping is crucial to avoid a dramatic drop in the sandwich effect caused by the steep negative gradient when $r$ is close to unity.

Ultimately, these effects remain uncertain due to ill-defined material properties, which must be further investigated experimentally.

%------
% TABLE
% -----
\begin{table}[!b]
\small
	\renewcommand{\arraystretch}{1.2} % Adjust table row spacing
    \setlength{\tabcolsep}{1.2mm} % Adjust table column spacing
	\caption{Samples $\mathrm{S}$ from a $\mathrm{L}_9$ orthogonal array and control samples $\mathrm{C}$.}
    \label{tab:taguchi_samples}
	\centering
	\begin{threeparttable}
	\begin{tabular}{cccr}
	Sample ID & Material & Infill (\%) & Lattice \\	
	\midrule
		S1 & HW-PLA & 20 & Gyroid \\
		S2 & HW-PLA & 30 & Honeycomb \\
		S3 & HW-PLA & 40 & Auxetic \\
		S4 & LW-PLA & 20 & Honeycomb \\
		S5 & LW-PLA & 30 & Auxetic \\
		S6 & LW-PLA & 40 & Gyroid \\
		S7 & ABS & 20 & Auxetic \\
		S8 & ABS & 30 & Gyroid \\
		S9 & ABS & 40 & Honeycomb \\
		C1 & Steel & - & - \\
		C2 & Aluminium & - & - \\
	\end{tabular}
\end{threeparttable}
\end{table}

% -----------------------
% EXPERIMENTAL VALIDATION
% -----------------------
\section{Experimental Validation}\label{sec:experimental_section}

The composite sheets and the 3D-printed cellular core were modeled with effective material properties. Those of the core, in particular, are difficult to pin down accurately. Building upon these modeling efforts, the following experiment aims to establish empirically the core best suited to haptic transparency.

\subsection{Design of Experiment Using the Taguchi Method}

The cellular cores were described experimentally by three independent variables, also called factors. They were chosen as the type of lattice, its material, and the infill percentage (the ratio of the actual volume of the lattice to that of its cuboidal envelope). Each factor was restricted to three different levels, listed in Tab.~\ref{tab:taguchi_samples}. These levels were picked according to manufacturing constraints and preliminary trials, which helped narrowing down the range of relevant values. A full factorial analysis, i.e. testing all $3^3$ combinations, would be too costly and time consuming. In his seminal work on off-line quality control~\cite{Taguchi-59, NairEtAl-92}, Taguchi proposed a method that greatly reduces the number of experiments required while keeping variance to a minimum. Applied to this case study, it provided a threefold reduction in the number of combinations, resulting in a well-balanced set of nine samples, labeled S1 to S9. The pairs of factors and levels that were retained followed Taguchi's $\mathrm{L_9}$ orthogonal array, shown in Tab.~\ref{tab:taguchi_samples}. This method uses quadratic loss functions to quantify the performance of each combination. They are usually linearized in the form of signal-to-noise ratios (S/N). In signal processing, the S/N is defined as the ratio of the signal power to the background noise power. In Taguchi's method, it is defined differently as the ratio of sensitivity to variability~\cite{TaguchiPhadke-89}. The bandwidth must be maximized while minimizing mass. In turn, the S/N formula for a response that must be maximized, also known as ``the-larger-the-better'' in relevant literature, is defined as,
\begin{equation}\label{eq:SNR_larger_better_bandwidth}
\mathrm{S/N} = -10\log \left(\frac{1}{n}\sum_{i=1}^{n}\frac{1}{{f_i}^2}\right)
\, ,
\end{equation}
where $n$ is the number of bandwidth measurements, $f_i$, for each combination. Conversely, a response that must be minimized, referred to as ``the-smaller-the-better'', is defined as, 
\begin{equation}\label{eq:SNR_smaller_better_mass}
\mathrm{S/N} = -10\log \left(\frac{1}{n}\sum_{i=1}^{n}{m_i}^2\right)
\, ,
\end{equation}
where $m_i$ is the mass of the sample in question. Although damping does not directly correlate to transparency, it remains a key property for smoothing out resonances and increasing bandwidth accordingly. It was estimated via the time constant, $\tau$, of an exponentially decaying impulse response. Quantifying transients unveils the capacity of lattices to dissipate energy. It calls for a S/N of the type ``the-smaller-the-better'', defined as,
\begin{equation}\label{eq:SNR_smaller_better_damping}
\mathrm{S/N} = -10\log \left(\frac{1}{n}\sum_{i=1}^{n}{\tau_i}^2\right)
\, .
\end{equation}

% ------
% FIGURE
% ------
\begin{figure*}[!b]
\centering
\includegraphics[width=181mm]{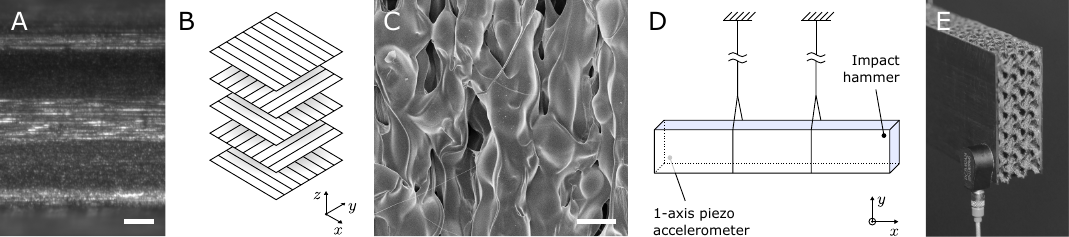}
\caption{(A) Cross-sectional microscopy of the carbon fiber face sheet in the $xz$-plane, taken with a Nikon SMZ800N. Fibers are revealed by dark-field illumination. The scale bar indicates \SI{100}{\micro\meter}. (B) Orthogonal layout of the 5-layer carbon fiber laminate. (C) Scanning electron microscope image of foaming LW-PLA, taken with a Zeiss Supra 40. The scale bar indicates \SI{100}{\micro\meter}. The sample was preliminary sputtered with a \SI{30}{\nano\meter}-thick layer of gold. (D) Schematic of the experimental setup used to measure the impulse response of a sample with free-free boundaries. (E) Photograph of the experimental setup.}
\label{fig:experimental}
\end{figure*}

\subsection{Face Sheet Manufacturing}

The face sheets were sized at $160\!\times\!30\!\times\!\SI{0.6}{\milli\meter\cubed}$ to meet the previously established dimensional requirements. They were cut out from a $\SI{0.6}{\milli\meter}$-thick ultra-high modulus carbon fiber panel (DragonPlate, Allred \& Associates) using a CNC router equipped with a diamond end mill, hard-coated via chemical vapor deposition (CVD). This panel was made from a 5-layer laminate of unidirectional, pitch-based, carbon fiber fabric pre-impregnated with epoxy (see cross-sectional micrograph in Fig.~\ref{fig:experimental}.A). Fabric layers were laid out following an orthogonal arrangement, purposefully aligned with the principal directions of the stress tensor, as shown in Fig.~\ref{fig:experimental}.B. It resulted in a density of $\SI[inter-unit-product = \!\cdot\!]{1.55}{\gram\per\centi\meter\cubed}$. Consistent manufacturing was confirmed by the low variance in face sheet masses $\SI[separate-uncertainty, multi-part-units=single]{4.55 \pm 0.06}{\gram}$ (mean $\pm$ std). A custom-made three-point bending test gave its flexural modulus, $\SI{294}{\mega\pascal}$ ($R^2>0.99$), beyond that of steel, $\SI{216}{\mega\pascal}$ ($R^2>0.99$). Excellent adhesion to the core is critical. Thus, the face sheets were bonded with a self-leveling epoxy adhesive (2217, 3M Company), which helped create a uniform bond. It was cured under pressure at \SI{60}{\celsius} for \SI{3}{\hour}.

\subsection{Core Manufacturing}

All lattices were designed with sufficiently smooth transitions along the $z$-axis to be self-supported. They were 3D-printed with a widely available printer (S3, Ultimaker). Fused deposition modeling (FDM) was preferred over higher-resolution options because it did not leave any chemical residue. This could have otherwise inhibited, or weakened, bonding with the outer faces. The spatial period of each lattice was individually tuned to match the required infill percentage. The polymeric filaments were extruded in $\SI{0.15}{\milli\meter}$-thick layers within the $xy$-plane. The intricate lattice features were printed with a wall thickness of $\SI{0.7}{\milli\meter}$, the smallest possible with the setup at hand. The polymers tested in this study were acrylonitrile butadiene styrene (ABS), with a density of $\SI[inter-unit-product = \!\cdot\!]{1.10}{\gram\per\centi\meter\cubed}$, and foaming polylactic acid (LW-PLA, ColorFabb). The latter behaves like regular PLA when extruded at \SI{200}{\celsius}, with a density of $\SI[inter-unit-product = \!\cdot\!]{1.20}{\gram\per\centi\meter\cubed}$. It was labeled HW-PLA. However, when extruded at \SI{250}{\celsius}, gas is released, which causes this material to foam. The resulting pores were about $\SI{100}{\micro\meter}$ in size, as revealed by the scanning electron microscopy (SEM) image in Fig.~\ref{fig:experimental}.C. The density of LW-PLA is $\SI[inter-unit-product = \!\cdot\!]{0.65}{\gram\per\centi\meter\cubed}$, about $54\%$ of that of HW-PLA.

\subsection{Acceleration Measurements and Processing}

The experiment was conducted with free-free boundary conditions, which were achieved by suspending the samples with two $\SI{1}{\meter}$-long, thin wires, as depicted in Fig.~\ref{fig:experimental}.D. To record the impulse responses, a monoaxial accelerometer (352A24, PCB Piezotronics) was fixed to the corner of the samples, hence avoiding nodal lines. It was attached by a thin layer of mounting wax (32279, Endevco) which provided good mechanical coupling. It measured out-of-plane accelerations with a $93.8\,\mathrm{mV\!\cdot\!g^{-1}}$ sensitivity and a $\SI{8}{\kilo\hertz}$ bandwidth, given for a $\pm 5\%$ error. Its light weight of $\SI{0.8}{\gram}$ accounted for $4\%$ of that of the lightest sample, thus imparting only minor inertial perturbations. The setup is pictured in Fig.~\ref{fig:experimental}.E.

An impact hammer (086E80, PCB Piezotronics) was used to create a mechanical impulse on the diagonally opposite corner, along the $z$-axis. Measurements were repeated 10 times per sample. The impact force was measured with a $22.5\,\mathrm{mV/N}$ sensitivity. It remained consistent throughout the experiment, at $2.60\pm\SI{0.83}{\newton}$. The impacts were about \SI{0.2}{\milli\second} long, which was sufficiently close to a Dirac. Trials with double tapping caused by surface rebound were rerun. A constant current of $\SI{4}{\milli\ampere}$ was fed to both the accelerometer and the hammer using a custom-made source built around a current regulator (LM317, Texas Instrument). Signals were digitized on 16 bits at a rate of $\SI{125}{\kilo\hertz}$ by an acquisition card (PCI-6121, National Instruments), followed by a zero-lag, 2-pole Butterworth band-pass filter with a \SI{50}{\hertz} cut-on and a \SI{10}{\kilo\hertz} cut-off.

Two control samples, labeled C1 and C2, were added to the experiment (see Tab.~\ref{tab:taguchi_samples}). They are solid blocks of steel and aluminium, respectively. They share the same dimensions than other samples, and thus provided ground truth modal responses.

%------
% TABLE
%------
\begin{table}[!h]
\small
	\renewcommand{\arraystretch}{1.2} % Adjust table row spacing
    \setlength{\tabcolsep}{1.5mm} % Adjust table column spacing
	\caption{Evaluation of performance metrics for haptic transparency.}
    \label{tab:taguchi_results}
	\centering
	\begin{threeparttable}
	\begin{tabular}{cccccccccc}		
	\multirow{3}[2]{*}{} & & \multicolumn{2}{c}{Bandwidth\,\textsuperscript{\textdagger} $f$} & & \multicolumn{2}{c}{Mass $m$} & & \multicolumn{2}{c}{Time constant\,\textsuperscript{\textdagger} $\tau$}\\ \cmidrule[0.4pt]{3-4}\cmidrule[0.4pt]{6-7}\cmidrule[0.4pt]{9-10}
    & & $\mu\pm\sigma$ & S/N & & $\mu$ & S/N & & $\mu\pm\sigma$ & S/N \\ [-0.5mm]
    & & (\SI{}{\hertz}) &(\SI{}{\deci\bel}) & & (\SI{}{\gram}) & (\SI{}{\deci\bel}) & & (\SI{}{\milli\second}) & (\SI{}{\deci\bel}) \\   		
    	\midrule[0.4pt]
		S1 & & $1309 \pm 52$  & 62.3 & & $25.0$ & -28.0 & & $4.4 \pm 0.5$ & 47.0\\
		S2 & & $1531 \pm 100$ & 63.6 & & $32.3$ & -30.2 & & $5.3 \pm 0.3$ & 45.5\\
		S3 & & $1556 \pm 58$  & 62.8 & & $37.7$ & -31.5 & & $4.6 \pm 0.5$ & 46.7\\
		S4 & & $600  \pm 34$  & 55.5 & & $19.1$ & -25.6 & & $6.0 \pm 0.7$ & 44.3\\
		S5 & & $800  \pm 32$  & 58.0 & & $21.1$ & -26.5 & & $6.2 \pm 0.4$ & 44.1\\
		S6 & & $681  \pm 36$  & 56.6 & & $27.2$ & -28.7 & & $5.8 \pm 0.5$ & 44.7\\
		S7 & & $1205 \pm 116$ & 61.5 & & $25.0$ & -28.0 & & $4.1 \pm 0.5$ & 47.6\\
		S8 & & $1270 \pm 55$  & 62.1 & & $32.2$ & -30.2 & & $3.0 \pm 0.5$ & 50.5\\
		S9 & & $1363 \pm 89$  & 62.6 & & $37.1$ & -31.4 & & $4.4 \pm 0.3$ & 47.2\\
		C1 & & $1398 \pm 29$  & 62.9 & & $474$  & -53.5 & & $768 \pm 6.9$ & 2.3\\
		C2 & & $1382 \pm 36$  & 62.8 & & $167$  & -44.5 & & $294 \pm 59$  & 10.5\\
	\end{tabular}
	\begin{tablenotes}
    \footnotesize
    \item \textsuperscript{\textdagger}\! Data were averaged over 10 trials (mean $\pm$ std).
    \end{tablenotes}
\end{threeparttable}
\end{table}

% ------
% FIGURE
% ------
\begin{figure*}[!t]
\centering
\includegraphics[width=181mm]{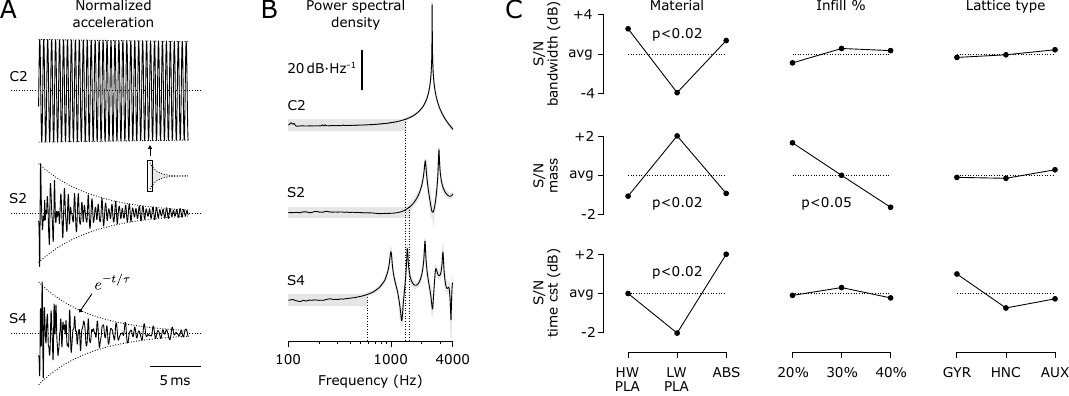}
\caption{(A) Underdamped impulse responses in the time domain for samples C2, S2, and S4, fitted with exponential envelopes in dotted lines. (B) Power spectral density with the bandwidth indicated by a shaded area. (C) Main effects associated to printing material, infill percentage, and lattice type, centered around their average values.}
\label{fig:main_effects}
\end{figure*}

% -------
% RESULTS
% -------
\subsection{Results}

The impulse responses in the time domain are presented in Fig.~\ref{fig:main_effects}.A for representative samples C2, S2, and S4. Aggregated results, including the remaining samples, are listed in Tab.~\ref{tab:taguchi_results}. The impulse responses were typical of underdamped oscillators, with an exponential decay that was fitted with an envelope in the form $e^{-t/\tau}$ using nonlinear least squares. The time constant, $\tau$, was averaged over ten trials. As expected, the metallic samples C1 and C2 exhibited minimal damping, characterized by a time constant two orders of magnitude greater than that of the polymeric sandwich structures.

The power spectral densities (PSD) shown in Fig.~\ref{fig:main_effects}.B were computed with Welch's method. Only spectra above \SI{100}{\hertz} were kept since the impact hammer method proved ineffective in exciting low frequencies. The bandwidth was defined at the $\SI[inter-unit-product = \cdot]{\pm3}{\deci\bel\per\hertz}$\! threshold, which corresponds to the shaded regions in Fig.~\ref{fig:main_effects}.B. High-performance sandwich samples, such as S2, had a superior bandwidth than the aluminium control C2 (about $+10\%$), for only a fraction of the mass (about $-81\%$). The first resonance of S2 was recorded at \SI{2160}{\hertz}, which was notably lower than that predicted by numerical simulations in Fig.~\ref{fig:dimensional_optim}.B. This confirms the need of experimental evaluation in dealing with ill-defined materials. In contrast to C1 and C2, the sandwich structures had a rich timbre, indicative of high-order dynamics which may have been caused by orthotropic lattice properties. All samples made from LW-PLA had significantly lower bandwidths. This drop is even more pronounced than what was predicted by theory (see sample S4 in Fig.~\ref{fig:material_optim}.B). All remaining samples met the \SI{1}{\kilo\hertz}-bandwidth requirement of haptic transparency.

\subsection{Main Effects and Analysis of Variance}

To develop effective design strategies, haptic researchers need to understand the relative significance of each factor (material, infill percentage, and lattice structure). A common approach is to quantify main effects, i.e. the impact of a factor on the goal, averaged across all levels of the other factors. For example, the main effects of the infill percentage are given by,
\begin{equation}\label{eq:mean_effects_1}
\left.\mathrm{S/N}\,\right\vert_{\mathrm{20\%}} = \frac{1}{3}\big(\!\left.\mathrm{S/N}\,\right\vert_{\mathrm{S1}}\!+\!\left.\mathrm{S/N}\,\right\vert_{\mathrm{S4}}\!+\!\left.\mathrm{S/N}\,\right\vert_{\mathrm{S7}}\!\big)
\, ,
\end{equation}
\begin{equation}\label{eq:mean_effects_2}
\left.\mathrm{S/N}\,\right\vert_{\mathrm{30\%}} = \frac{1}{3}\big(\!\left.\mathrm{S/N}\,\right\vert_{\mathrm{S2}}\!+\!\left.\mathrm{S/N}\,\right\vert_{\mathrm{S5}}\!+\!\left.\mathrm{S/N}\,\right\vert_{\mathrm{S8}}\!\big)
\, ,
\end{equation}
\begin{equation}\label{eq:mean_effects_3}
\left.\mathrm{S/N}\,\right\vert_{\mathrm{40\%}} = \frac{1}{3}\big(\!\left.\mathrm{S/N}\,\right\vert_{\mathrm{S3}}\!+\!\left.\mathrm{S/N}\,\right\vert_{\mathrm{S6}}\!+\!\left.\mathrm{S/N}\,\right\vert_{\mathrm{S9}}\!\big)
\, .
\end{equation}
The remaining main effects were found similarly. They are shown in Fig.~\ref{fig:main_effects}.C, centered around their average. The contribution of each factor was derived from a one-way analysis of variance (ANOVA). Taguchi experiments suffer from a fully saturated factorial design. Instead, an unsaturated description was obtained by discarding the factor with the lowest mean square effect (MSE), which thus served as the mean square error. Statistical results are reported in Tab.~\ref{tab:taguchi_anova}, including the degrees of freedom (DoF), the sum of squares (SS), the mean squares (MS), the F-value (Fisher test), the p-value, and the contribution percentage.

%------
% TABLE
%------
\begin{table}[!b]
\small
	\renewcommand{\arraystretch}{1.2} % Adjust table row spacing
    \setlength{\tabcolsep}{1.2mm} % Adjust table column spacing
	\caption{One-way ANOVA on the results obtained via Taguchi's method.}
    \label{tab:taguchi_anova}
	\centering
	\begin{threeparttable}
	\begin{tabular}{clcccccc}
	Goal & Factor & DoF & SS & MS & F-value & p-value\,\textsuperscript{\textdagger} & Cont. (\%)\\	
	\midrule[0.4pt]
	    \multirow{3}{*}{$f$}
    		& Material & 2 & 72.5 & 36.2 & 76.1 & 0.013 & 94 \\
	    & Infill & 2 & 3.8 & 1.9 & 1.0 & 0.500 & 5 \\
	    & Lattice & 2 & 1.0 & 0.5 & - & - & 1 \\
		\midrule[0.4pt]
	    \multirow{3}{*}{$m$}		    
		& Material & 2 & 18.1 & 9.1 & 52.9 & 0.019 & 52 \\
    		& Infill & 2 & 16.3 & 8.1 & 47.4 & 0.021 & 47 \\
    		& Lattice & 2 & 0.3 & 0.2 & - & - & 1 \\	
		\midrule[0.4pt]    
	    \multirow{3}{*}{$\tau$} 
		& Material & 2 & 24.3 & 12.2 & 53.3 & 0.018 & 82 \\    		
		& Infill & 2 & 0.5 & 0.2 & - & - & 2 \\	   
		& Lattice & 2 & 4.8 & 2.4 & 10.6 & 0.086 & 16 \\	
		\end{tabular}
	\begin{tablenotes}
    \footnotesize
    \item \textsuperscript{\textdagger}\! Computed at a 0.05 significance level.
    \end{tablenotes}
\end{threeparttable}
\end{table}

The type of lattice had no significant effect on either the bandwidth or the mass. Although the gyroid lattice imparted slightly lower time constants, i.e. a higher S/N, these differences were not statistically significant (p$\,>\,$0.05). This suggests that a gyroid core may enhance damping to some extent, though the exact mechanism remains unknown. Essentially, haptic designers can freely choose from different cellular geometries based on other considerations.

As shown in Fig.~\ref{fig:main_effects}.C, the S/N of the mass is proportional to the infill percentage (p$\,<\,$0.05). While it may be trivial since the infill represents the volume of matter, this provides further validation that the samples were produced to specifications.

% ------
% FIGURE
% ------
\begin{figure*}[!t]
\centering
\includegraphics[width=181mm]{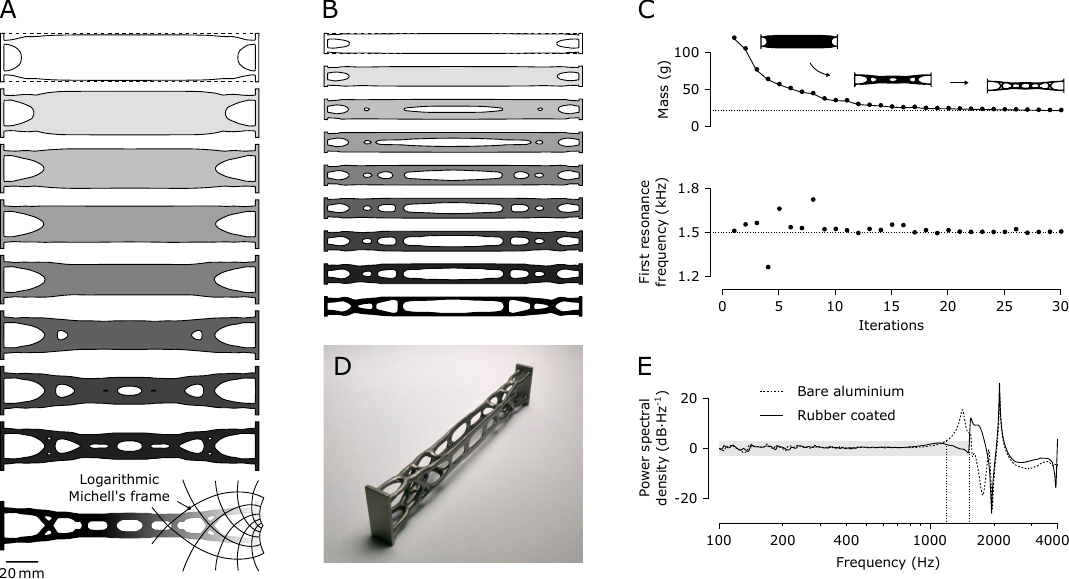}
\caption{(A) Evolution of the geometry during the optimization process (side view), starting from a cuboid delineated by the dashed lines. An overlay of logarithmic Michell's frame coincides with the optimized topology. (B) Corresponding top view of the optimization process. (C) Mass and first resonance frequency at different stages of the optimization. (D) Photograph of the 3D-printed aluminium prototype before rubber coating. (E) Power spectral density of the impulse response measured for an aluminium prototype, either bare or rubber coated.}
\label{fig:topology_optim}
\end{figure*}

Materials used for 3D printing had a significant impact an all goals. They were responsible for $94\%$ of the impact on bandwidth, compared to other factors (p$\,<\,$0.02). A similar trend was observed for the time constant (p$\,<\,$0.02), which supports the correlation between damping and bandwidth. In fact, despite a first resonance frequency greater than S2, metallic sample C2 had a lower bandwidth due to a lack of damping, as shown in Fig.~\ref{fig:main_effects}.B. Surprisingly, LW-PLA performed significantly worse than other materials, hampered by higher time constants. Its foam-like structure, fragmented by submillimetric air pockets (see Fig.~\ref{fig:experimental}.A) was instead expected to increase wave absorption. This may also be attributed to a non-linear modulus reduction mechanism caused by porosities.

Overall, the best combination consists of an arbitrary lattice made from a $20\%$ to $30\%$ infill of either ABS or HW-PLA. Such cellular core, sandwiched between UHM-CFRP face sheets, provided unprecedented haptic transparency with a flat spectrum safely above \SI{1}{\kilo\hertz} and an effective density of only about $\SI[inter-unit-product = \cdot]{0.5}{\gram\per\centi\meter\cubed}$ (see Tab.~\ref{tab:taguchi_results}). In essence, it is a scalable and easy-to-manufacture metamaterial, endowed with properties that surpass those of natural materials.

% -------------------
% ALTERNATIVE DESIGNS
% -------------------
\section{Alternative Designs}

The previously examined design paradigm provided an exceptional level of haptic transparency. With current technological advancements, it prompts the question of whether there is an alternative approach that can rival this paradigm in terms of optimality. Although a comprehensive answer could not be provided, another promising concept was investigated as a point of comparison.

\subsection{Generative Design and Topology Optimization}

In particular, topology optimization has received growing interest in the generative design framework~\cite{SigmundMaute-13}. Structures obtained with such methods often look borrowed from nature. They are characterized by organic and highly convoluted features that would otherwise exceed one's creative abilities. These advances go in pair with additive manufacturing, whose technological maturity has made practical realizations possible. Since the pioneering work of Michell in 1904~\cite{Michell-04} and key improvements, for example those of Rozvany in 1992~\cite{RozvanyEtAl-92}, numerous algorithms have emerged and numerical formulations have made their way into commercially available software.

\subsection{Numerical Implementation}

In order to compare generative design to the aforementioned composite structures, a topology optimization study was carried out using the software PTC Creo 9.0. The goal was set to minimize the mass while maintaining a first resonance frequency above \SI{1.5}{\kilo\hertz}. This threshold was chosen to compensate for a high quality factor, which was expected to reduce the bandwidth. For a fair comparison, the design space was set to a cuboid with dimensions identical to those of previous samples S1 to S9. A $\SI{2}{\milli\meter}$-thick region taken from its extremities was set to not be modified. Without this condition, the solver would otherwise converge towards an empty geometry. The design space was further constrained to three orthogonal symmetry planes. The model was assumed to be made out of aluminium, described by a Young's modulus $E = \SI{70}{\giga\pascal}$, a Poisson's ratio $\nu = 0.33$, and a density $\SI[inter-unit-product = \!\cdot\!]{2.7}{\gram\per\centi\meter\cubed}$.

The computed geometry is depicted at different stages of the optimization process in Fig.~\ref{fig:topology_optim}.A (side view) and Fig.~\ref{fig:topology_optim}.B (top view). Interestingly, matter ended up being pushed away from the neutral axes, probably in an effort to increase flexural rigidity. Even with a radically different approach, the fundamentals of sandwich materials were recovered. In addition, the optimized part possessed topological features reminiscent of Michell's frames, for instance the logarithmic spirals drawn in Fig.~\ref{fig:topology_optim}.A. As shown in Fig.~\ref{fig:topology_optim}.C, the optimization process converged towards a minimal mass of \SI{21.6}{\gram} after 30 iterations, while satisfying the frequency threshold of \SI{1.5}{\kilo\hertz}. A post-hoc finite element analysis revealed the modal deformation. The first mode involved a combination of bending and torsion, confirming optimal material usage to meet the frequency thresholds for both dynamic loads.

\subsection{Experimental Validation}

The optimized topology was 3D-printed in an aluminium alloy ($\mathrm{AlSi_{10}Mg}$) using selective laser melting (SLM). The print was further glass bead blasted to remove residual manufacturing defects. The prototype is pictured in Fig.~\ref{fig:topology_optim}.D. The impulse response was measured using the experimental setup previously described. The power spectral density, given in Fig.~\ref{fig:topology_optim}.E, revealed a first resonance at \SI{1417}{\hertz}, which was in close agreement with numerical simulations. Despite being made of a metallic alloy, this structure had a lower quality factor than that of samples C1 and C2. This could be attributed the porous microstructure created by locally melted powder. Nonetheless, the usable bandwidth was lowered down to $1181\pm\SI{45}{\hertz}$ for a sample mass of \SI{20.0}{\gram}. To enhance damping, a synthetic rubber coating with a Shore A 70 hardness (Plasti Dip) was sprayed onto the prototype in three layers. As shown in Fig.~\ref{fig:topology_optim}.E, it altered the shape of the first resonance, resulting in an increased bandwidth of $1516\pm\SI{19}{\hertz}$, at the cost of a minimal mass increase of \SI{0.3}{\gram} only.

This alternative concept demonstrated slightly better, yet comparable performance to the sandwich structure paradigm. Specifically, it achieved a $16\%$ gain in frequency with a $19\%$ mass reduction compared to the sample S1, arguably among the best. However, it did lack design freedom and commonly available manufacturing options. The convergence towards a technological limit can be inferred from the similarity between both solutions.

% ----------
% CONCLUSION
% ----------
\section{Conclusion}

This work introduced a new design paradigm to help haptic engineers create highly transparent tools that do not taint haptic feel. It consists of a sandwich structure made from ultra-high modulus carbon fiber sheets arranged around an optimized 3D-printed lattice. Through theoretical, numerical, and experimental studies, it was demonstrated that this approach offers both increased bandwidth and reduced inertial mass. In that regard, the printing material was found to be the most critical property of the cellular cores. Normalized samples and standardized test methods were used to establish an easily replicable benchmark for other haptic researchers to compare their own approaches.

Topology optimization was investigated as an alternative. Interestingly, it converged on principles analogous to sandwich structures. Both techniques performed equally well despite the use of different cutting-edge materials and fabrication methods. This indicates that the unprecedented transparency shown in this study may have reached a technological limit. Topology optimization unlocks a wide range of 3D design possibilities, which could be combined with sandwich structures, better suited for 2D, to create more versatile haptic devices.

While this medium primarily targets haptic instruments and interfaces, it also has the potential to enhance the dynamic response of robotic manipulators. In addition to its engineering utility, it provides opportunities to create tactual curiosities and illusions. Commonplace objects often lack haptic transparency, which our somatosensory system has adapted to. In turn, making replicas out of high-performance sandwich composites could present intriguing case studies, beyond the realm of natural interactions. These could be used to address unresolved questions related to touch projection and the hypothetical existence of a tactile horizon.

% -----------------
% DATA AVAILABILITY
% -----------------
\section*{Data Availability}

The data supporting the findings of this study are available in the Zenodo open repository at \url{https://doi.org/10.5281/zenodo.13305665}.

% ---------------
% ACKNOWLEDGMENTS
% ---------------
\section*{Acknowledgments}

The authors would like to thank the Institut des NanoSciences de Paris (INSP) for granting access to the scanning electron microscope.

% ------------
% BIBLIOGRAPHY
% ------------
\bibliographystyle{IEEEtran}
\bibliography{2024_TD_SH_VH.bbl}

% ---
% END
% ---
\end{document}